\documentclass[letterpaper]{article}
\usepackage{aaai}
\usepackage{times}
\usepackage{helvet}
\usepackage{courier}
\usepackage{amsmath}
\usepackage[utf8]{inputenc}
\usepackage{graphicx}
\begin{document}
%
\title{A Review of Repository Level Prompting for LLMs}
\author{ Douglas Schonholtz \\
Northeastern University, Boston, Masssachusetts, 02115\\
schonholtz.d@northeastern.edu \\
}
\maketitle
\begin{abstract}
As coding challenges become more complex, recent advancements in Large Language Models (LLMs) have led to notable successes, such as achieving a 94.6\% solve rate on the HumanEval benchmark. Concurrently, there is an increasing commercial push for repository-level inline code completion tools, such as GitHub Copilot and Tab Nine, aimed at enhancing developer productivity. This paper delves into the transition from individual coding problems to repository-scale solutions, presenting a thorough review of the current literature on effective LLM prompting for code generation at the repository level. We examine approaches that will work with black-box LLMs such that they will be useful and applicable to commercial use cases, and their applicability in interpreting code at a repository scale. We juxtapose the Repository-Level Prompt Generation technique with RepoCoder, an iterative retrieval and generation method, to highlight the trade-offs inherent in each approach and to establish best practices for their application in cutting-edge coding benchmarks. The interplay between iterative refinement of prompts and the development of advanced retrieval systems forms the core of our discussion, offering a pathway to significantly improve LLM performance in code generation tasks. Insights from this study not only guide the application of these methods but also chart a course for future research to integrate such techniques into broader software engineering contexts.
\end{abstract}

\section{Introduction}

The advent of Large Language Models (LLMs) has revolutionized the field of automated code generation, offering solutions that range from simple code snippets to complex blocks that fit into larger systems. The HumanEval benchmark, which emerged as a standard for evaluating the capabilities of these models, has shown that LLMs can achieve high success rates, often outperforming traditional programming approaches \cite{chen2021codex}. However, as we shift from isolated challenges to the nuanced realm of repository-level development, new complexities arise. Developers don't just need code that works; they need code that integrates seamlessly with existing codebases—a challenge that requires a deep understanding of context, dependencies, and the idiosyncrasies of each repository.

The concept of Retrieval Augmented Generation (RAG) presents a paradigm shift in this landscape. By leveraging existing code in a repository as a knowledge base, RAG enables LLMs to generate more accurate and contextually relevant code completions. This process is akin to a seasoned developer recalling similar problems and solutions from past experiences to tackle current coding challenges.

Furthering this idea, the integration of another LLM for the purpose of prompt generation opens up new avenues for contextual understanding. It allows for a more dynamic interaction with the codebase, where the LLM not only generates code but also assists in identifying the most relevant pieces of information to guide its generation process. This meta-level operation stands to significantly enhance the model's performance by refining its focus and adapting its output to the specific context of the repository in question.

This paper provides an in-depth analysis of these innovative approaches, particularly focusing on their effectiveness at the repository level—a scale that presents a more realistic and challenging scenario for LLMs. We explore two pioneering methods: the Repository-Level Prompt Generation technique and RepoCoder, an iterative retrieval and generation method. By juxtaposing these methodologies, we aim to uncover the trade-offs they present and establish a set of best practices for leveraging LLMs in repository-level code generation. Our review does not merely highlight their potential but also underscores the necessity for sophisticated retrieval systems that can keep pace with the evolving complexity of software development tasks.

As we navigate through these methodologies, our narrative will unravel the intricate tapestry of LLM capabilities, their limitations, and the prospects of combining various strategies to create a more potent tool for developers. The synthesis of iterative refinement of prompts with advanced retrieval systems promises a future where LLMs could become indispensable assistants to programmers, offering intelligent suggestions that streamline the development process and elevate the quality of software engineering.

\cite{chen2021codex}, \cite{zhang2023repocoder}, \cite{shrivastava2023repositorylevel}

\pagebreak

\section{Repository Level Retrieval Techniques}

In order to be able to make good code completions at the repository level, an LLM must be prompted with relevant context so that it can infer what code it should generate.

\begin{figure}[ht]
  \centering
  \includegraphics[width=0.9\linewidth]{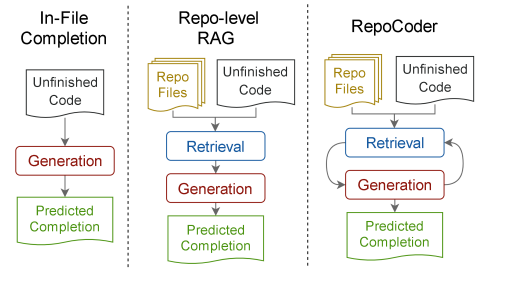}
  \caption{The iterative process of the RepoCoder framework. As seen in RepoCoder \cite{zhang2023repocoder}}
  \label{fig:repocoder}
\end{figure}

RepoCoder\cite{zhang2023repocoder} uses BM25\cite{Amati2009}, a sparse retrieval technique, across sliding windows of code to retrieve code relevant to the the generated code, and the code around the hole the generated code fills. Then that generation is used as a query to find other relevant code with the same BM25/bag of words retrieval system. This can then be done iteratively potentially improving every time this process is run. This process is illustrated in figure \ref{fig:repocoder}

The generated code can be modeled with this equation: \(\hat{Y}^{i} = M(C_{\text{ret}}^{i}, X).\). Where Y is the output generated code to fill the missing gap, X is the surrounding code, and $C_{repo}$ is the repository context which is initially empty.

The context retrieval process can be modeled with this equation.
\(C_{\text{ret}}^{i} = R(C_{\text{repo}}, X, \hat{Y}^{i-1}).\)
Where $C_{\text{ret}}^{i}$ is iteration i of context retrieval. 

In contrast, Repo Level Prompting presents a significantly more complicated approach as it's paper focuses on optimizing retrieval instead of optimizing the iteration process with an existing retrieval method.

The repository-level Prompt Generation paper compiles code prompts from various relevant code segments, which are then processed by the OpenAI Codex model. The effectiveness of each prompt is assessed based on whether it successfully completes the code gaps, marked by a binary score indicating success or failure. All of the prompting options are indexed into a vector, for a given repository completion the ground truth vector is defined as a one in each index of that vector where the associated prompt resulted in a successful generation. \cite{shrivastava2023repositorylevel}.

This can then be used to build a loss function which will allow classification of these prompts as useful or not in arbitrary code contexts. The loss function for the model is an average of the individual losses across code holes, formalized in equation \ref{eq:PromptGenLoss}.

\begin{equation}
\mathcal{L} = \frac{1}{N} \sum_{h=1}^{N} \mathcal{L}^h = \frac{1}{N} \sum_{h=1}^{N} \frac{1}{M^h} \sum_{p=1}^{M^h} BCE(\hat{y}_p^h, y_p) \times T_p^h
\label{eq:PromptGenLoss}
\end{equation}

where \( M^h = \sum_p r_p^h \) represents the total number of pertinent prompt proposals for each code gap \( h \), \( N \) denotes the aggregate number of code gaps during training, and \( BCE \) signifies the binary cross-entropy loss \cite{shrivastava2023repositorylevel}.

To predict optimal prompts for code generation by LLMs, two models are constructed: RLPG-H, a two-layer dense network featuring a ReLU non-linearity, and RLPG-R, which utilizes a single multi-head attention layer. Both of these models are placed on top of frozen embedding models and then trained to attempt to classify the optimal prompts to choose given the missing code context. Despite RLPG-R using multi-head attention, it is less performant than the dense model and it underperforms compared to RLPG-H.

The prompts generated are based off of the current, parent, children, imported, fuzzy matched, and sibling classes, methods, and files. Then these chunks of context are added above the default completion in no particular order. They also are truncated as necessary from the front or back to fit into the context window.

\section{Datasets}

The datasets used were different. RepoCoder uses a collection of repositories they extracted via the GitHub API as seen in Table \ref{tab:RepoCoderData}.

\begin{table}[ht]
\centering
\caption{Function Body Completion Dataset and Line and API Invocation Completion Datasets. F is the total number of python files, while N is the number of extracted Samples.}
\begin{tabular}{|c|l|l|l|c|c|}
\hline 
\textbf{Name} & \textbf{License} & \textbf{Created} & \textbf{F.} & \textbf{N.} \\ \hline
imagen & MIT License & 2022-05-23 & 14 & 67 \\
tracr & Apache V2.0 & 2022-12-01 & 56 & 146 \\
lightmmmm & Apache V2.0 & 2022-10-10 & 36 & 64 \\
inspection & Apache V2.0 & 2022-05-05 & 16 & 32 \\
omnivore & CC BY-NC 4.0 & 2022-01-20 & 66 & 22 \\
redframes & BSD-2-Clause & 2022-08-21 & 49 & 42 \\ \hline
\multicolumn{5}{|l|}{Line and API Invocation Completion Datasets} \\ \hline
rl & MIT License & 2022-02-01 & 165 & 400 \\
ACE & Apache V2.0 & 2022-11-23 & 425 & 400 \\
vizier & Apache V2.0 & 2022-02-16 & 188 & 400 \\
fortuna & Apache V2.0 & 2022-11-17 & 168 & 400 \\
evaluate & Apache V2.0 & 2022-03-30 & 180 & 400 \\
diffusers & Apache V2.0 & 2022-05-30 & 305 & 400 \\
nerfstudio & Apache V2.0 & 2022-05-31 & 357 & 400 \\
FedScope & Apache V2.0 & 2022-03-24 & 443 & 400 \\ \hline
\end{tabular}
\label{tab:RepoCoderData}
\end{table}

\begin{figure*}[ht]
  \centering
  \includegraphics[width=0.9\linewidth]{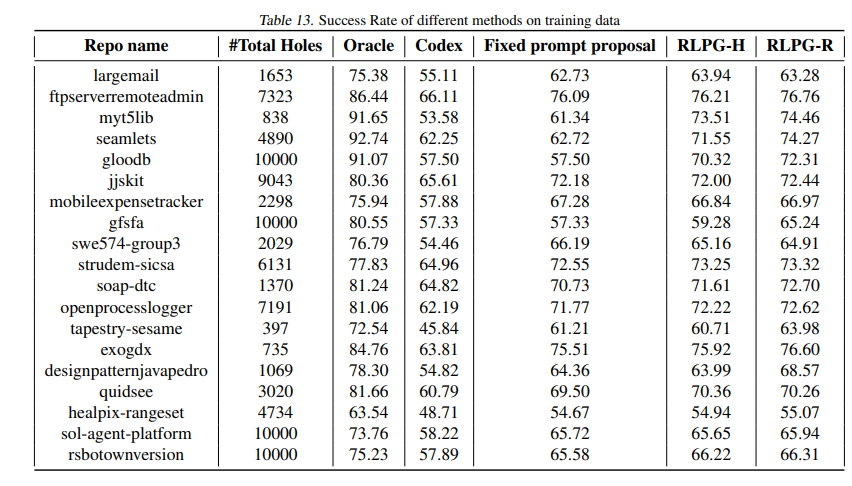}
  \caption{Repositories used in RepoLevel Prompting}
  \label{fig:RepoLevelRepos}
\end{figure*}

Meanwhile, Repo Level Prompting used repositories from Google as they feared that GitHub's data was used for training OpenAI's Codex model that they were testing. These repositories can be found in Figure \ref{fig:RepoLevelRepos}. This is a collection of Java repositories. It is worth noting, that the dataset for repo level prompting is far larger and more comprehensive, as the number of total holes there is much larger than the number of samples found in the number of samples extracted from RepoCoder.

\begin{table}[ht]
\centering
\caption{Repo Level Prompting Dataset}
\begin{tabular}{lcccc}
\hline
\textbf{Feature} & \textbf{Train} & \textbf{Val} & \textbf{Test} & \textbf{Total} \\ \hline
\# Repositories & 19 & 14 & 14 & 47 \\
\# Files & 2655 & 1060 & 1308 & 4757 \\
\# Holes & 92721 & 48548 & 48288 & 189557 \\ 
\hline
\end{tabular}
\label{tab:dataset_for_repo_level_prompting}
\end{table}

For Repo Level Prompting you can also see the total number of holes and breakdown by train, validation and test splits for all repositories in Table \ref{tab:dataset_for_repo_level_prompting}

\section{Comparing Oracle Retrieval Systems}

Both RepoCoder, and RepoLevel Prompting compare themselves to optimal oracle retrieval systems. In RepoCoder, this means doing the retrieval based on the ground truth code to fill the hole that the LLM normally would first attempt to hallucinate before doing it's first round of retrieval. In RepoLevel prompting, this means for their vector Y of possible successful prompts, using that set of optimal working prompts to search for code. In both cases, this creates an upper bound showing the limit of capability of these methods.

\begin{table}[ht]
\centering
\caption{Performance of the oracle relative to Codex for repo level prompting.}
\begin{tabular}{lccc}
\hline
\textbf{Split} & \textbf{SR Codex(\%)} & \textbf{SR Oracle(\%)} & \textbf{Rel. \textuparrow Codex(\%)} \\ \hline
Train & 59.78 & 80.29 & 34.31 \\
Val & 62.10 & 79.05 & 27.28 \\
Test & 58.73 & 79.63 & 35.58 \\ \hline
\end{tabular}
\label{tab:performance_oracle_repo_level_prompting}
\end{table}

You can see the effectiveness of the oracle and the relative improvement for it in Repo Level Prompting in Table \ref{tab:performance_oracle_repo_level_prompting}. You can find a breakdown of how the oracle does for RepoCoder in a later set of results tables. The most important take away from these though is that the oracle shows a potential upper bound with the codex model for Repo Level Prompting of 35.48\% on the test set. While RepoCoder shows up to a 90\% relative performance improvement on some function completion tests and a 45\% improvement on the exact match benchmark with GPT-3.5 when compared to in-file prompting. These methods will be expanded on more in the Validation of Code Generation section where we cover results.

\section{Locations of Text in Repositories for Optimal Retrieval}

One of the crucial findings of both RepoCoder and Repo Level Prompting is where the critical code is stored. This information can be used to build future potentially more effective retrieval systems to power these code generating LLMs.

RepoCoder's breakdown in Figure \ref{fig:repocoderLocationStats} shows that fuzzy import and names are extremely valuable with usage marking in the 50 and 80 percentiles for cases where the repocoder method beat the naive in file completion method.

\begin{figure}[ht]
  \centering
  \includegraphics[width=0.9\linewidth]{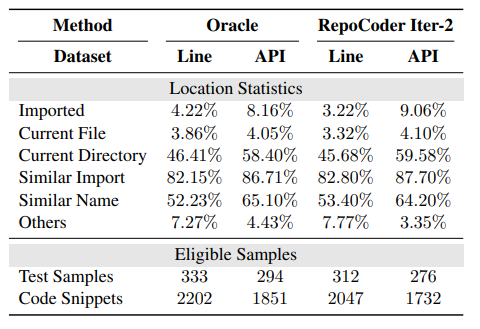}
  \caption{Locations of retrieved code snippets when
the Oracle/RepoCoder method outperforms the In-File
completion method using GPT-3.5-Turbo on the line
and API completion datasets}
  \label{fig:repocoderLocationStats}
\end{figure}

We can contrast the findings of RepoCoder with those of Repo Level Prompting in Figure \ref{fig:repocoderLocationStats} and in Figure \ref{fig:broadLocationStats}.

\begin{figure}[ht]
  \centering
  \includegraphics[width=0.9\linewidth]{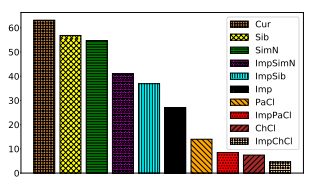}
  \caption{Mean success rates
of different prompt sources when they are applicable.
}
  \label{fig:repocoderLocationStats}
\end{figure}

\begin{figure*}[ht]
  \centering
  \includegraphics[width=0.9\linewidth]{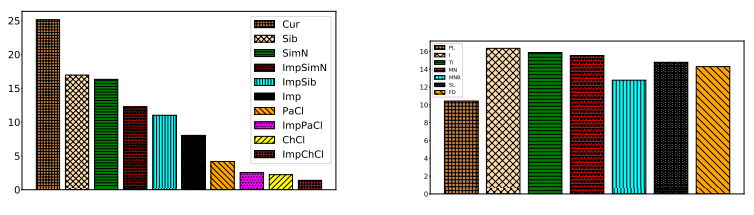}
  \caption{(Left) Normalized success rate of prompt sources when applicable, (Right) Normalized success rate of prompt context types
when applicable
}
  \label{fig:broadLocationStats}
\end{figure*}

\subsection{Short Hand File Source References}

To understand these references we also have to understand these short hand labels. They are broken out for convenience below.

\begin{itemize}
    \item \textbf{Current}: take code from the current file excluding the contents of the target hole. The current file is the file that contains the target hole. The code in the current file (e.g. the lines after the hole position) can be very useful in predicting the target hole.
    \item \textbf{Parent Class}: take code from the file that contains the parent of the class to which the target hole belongs. The intuition behind this is to account for cases where a method present in the parent class is invoked in the current file (i.e. the child class).
    \item \textbf{Import}: take code from the import files used in the current file. The dependencies specified via imports can provide useful cues to predict the target hole.
    \item \textbf{Sibling}: take code from the files that are in the same directory as the current file. Files in the same directory tend to share code variables (e.g. identifiers).
    \item \textbf{Similar Name}: take code from files that have a similar name as the current file. Similar names are determined by splitting the file name based on underscore or camel-case formatting and then matching parts of the filename. If one or more parts matches, two files are considered to have similar names. The intuition behind this is that software developers tend to name files based on the functionality of the code written in that file. Therefore, a similar name file might contain some portion of the code that is common with the current file and hence might be useful for predicting the target hole.
    \item \textbf{Child Class}: take code from files that have the current file as their parent class file.
    \item \textbf{Import of Parent Class}: take code from the import files used in the parent class files.
    \item \textbf{Import of Sibling}: take code from the import files used in the sibling files.
    \item \textbf{Import of Similar Name}: take code from the import files used in the similar name files.
    \item \textbf{Import of Child Class}: take code from the import files used in the child class files.
\end{itemize}

\subsection{Short Hand Data to be Extracted from Files}
To understand, the figure to the right in Figure 5, we have to understand this corresponding list as well.

\begin{itemize}
    \item \textbf{Post Lines (PL)}: Take all the lines after the target hole line till the end of the current file.
    \item \textbf{Identifiers (I)}: Take all the identifiers used in the prompt source.
    \item \textbf{Type Identifiers (TI)}: Take all the type identifiers used in the prompt source.
    \item \textbf{Field Declarations (FD)}: Take all the field declarations used in the prompt source.
    \item \textbf{String Literals (SL)}: Take all the string literals used in the prompt source.
    \item \textbf{Method Names (MN)}: Take all the method names along with their signatures used in the prompt source.
    \item \textbf{Method Names and Bodies (MNB)}: Take all the method names along with their signatures and corresponding bodies used in the prompt source.
\end{itemize}

\section{Validation of Code Generation}
RepoCoder evaluates their system on three separate benchmarks they build. The first benchmark was line completion, completing a single line in a repository. The second benchmark was function completion, where the model has to complete a function which has had some subset of its lines removed. The results for this benchmark can be seen in Table \ref{tab:line_completion_repocoder}.

\begin{table}[ht]
\centering
\caption{Line Completion performance evaluation using RepoCoder for exact match (EM) and Normalized Edit Distance (NED)}
\begin{tabular}{lcccccc}
\hline
\textbf{Metric} & \textbf{Oracle} & \textbf{In-File} & \multicolumn{4}{c}{\textbf{RepoCoder Iterations}} \\
 & & & \textbf{1} & \textbf{2} & \textbf{3} & \textbf{4} \\ \hline
\multicolumn{7}{c}{GPT-3.5-Turbo} \\
EM & 57.75 & 40.56 & 55.31 & 56.81 & 57.00 & 56.63 \\
NED & 75.43 & 65.06 & 74.38 & 75.11 & 75.30 & 75.10 \\ \hline
\multicolumn{7}{c}{CodeGen-Mono-6B} \\
EM & 48.81 & 34.56 & 45.81 & 47.06 & 47.75 & 47.44 \\
NED & 71.02 & 60.67 & 69.21 & 70.10 & 70.73 & 70.19 \\ \hline
\multicolumn{7}{c}{CodeGen-Mono-2B} \\
EM & 47.31 & 33.63 & 44.56 & 46.94 & 46.69 & 47.13 \\
NED & 69.80 & 58.99 & 67.68 & 68.82 & 68.62 & 68.92 \\ \hline
\multicolumn{7}{c}{CodeGen-Mono-350M} \\
EM & 45.19 & 29.56 & 41.88 & 43.06 & 43.94 & 43.06 \\
NED & 67.20 & 55.39 & 65.05 & 65.66 & 65.97 & 65.62 \\ \hline
\end{tabular}
\label{tab:line_completion_repocoder}
\end{table}

The third benchmark was API invocation completion. This included 1046 and 1086 samples that were run on GPT-3.5 Turbo and other smaller CodeGen models respectively. Where an api invocation was just an api that was invoked somewhere in the codebase and that was the code that had to be completed. All of these methods are then evaluated against an exact match metric, a metric to determine if the generated code matches character for character with the true expected code. They are then also tested with a normalized edit distance metric (NED), which is the Levenshtein distance\cite{Levenshtein1965BinaryCC} divided by the maximum character count between the generated code and the ground truth. You can see the results for this measurement in Table \ref{tab:api_invocation_completion_repocoder}.

\begin{table}[ht]
\centering
\caption{API Invocation Completion performance evaluation using RepoCoder.}
\begin{tabular}{lcccccc}
\hline
\textbf{Metric} & \textbf{Oracle} & \textbf{In-File} & \multicolumn{4}{c}{\textbf{RepoCoder Iterations}} \\
 & & & \textbf{1} & \textbf{2} & \textbf{3} & \textbf{4} \\ \hline
\multicolumn{7}{c}{GPT-3.5-Turbo} \\
EM & 50.13 & 34.06 & 47.69 & 49.19 & 49.44 & 49.56 \\
NED & 74.50 & 63.22 & 73.63 & 74.43 & 74.59 & 74.48 \\ \hline
\multicolumn{7}{c}{CodeGen-Mono-6B} \\
EM & 40.25 & 26.19 & 36.69 & 38.88 & 39.13 & 39.31 \\
NED & 67.94 & 56.45 & 64.20 & 65.52 & 65.53 & 65.90 \\ \hline
\multicolumn{7}{c}{CodeGen-Mono-2B} \\
EM & 39.44 & 25.44 & 35.44 & 37.56 & 38.44 & 38.25 \\
NED & 66.78 & 56.88 & 63.47 & 64.15 & 64.53 & 64.60 \\ \hline
\multicolumn{7}{c}{CodeGen-Mono-350M} \\
EM & 34.88 & 22.19 & 31.75 & 33.88 & 33.75 & 33.81 \\
NED & 63.06 & 52.24 & 59.82 & 61.03 & 60.96 & 61.06 \\ \hline
\end{tabular}
\label{tab:api_invocation_completion_repocoder}
\end{table}

Both of these results rely heavily on the exact match and NED metric. Such evaluations, however, are limited by their syntactic focus, neglecting the semantic similarity vital for grasping the intended code functionality. Moreover, the approach does not consider anonymized variables common in arbitrary solutions, nor does it acknowledge various valid problem-solving methods.

To mitigate these limitations, test case validation is also employed, though the methodology lacks a pre-and post-modification test state analysis, raising concerns about the integrity of the test cases and their coverage of the modified code. Meaning that verification that the tests pass before and after the code is implemented in the ground truth state, and that the unchanged code causes the modified tests to fail, as is described in other papers.\cite{jimenez2023swebench}. Despite these concerns, the pass rate of 44.6\% in repository test cases reported by RepoCoder is promising, especially when juxtaposed with a mere 23.32\% achievement through in-file retrieval. The improvements of 8\% and 10\% in exact and Levenshtein match scores over the baseline model further underscore the potential of the repo coder methodology. These results are shown in Table \ref{tab:repocoder_testing}

\begin{table}[ht]
\centering
\caption{RepoCoder Pass Rate against unit tests.}
\begin{tabular}{|c|c|c|c|c|c|c|}
\hline
\textbf{N.} & \textbf{Oracle} & \textbf{In-File} & \multicolumn{4}{c|}{\textbf{RepoCoder Iterations}} \\
& & & \textbf{1} & \textbf{2} & \textbf{3} & \textbf{4} \\ 
 \hline
67 & 56.72 & 29.85 & 53.73 & 55.22 & 55.22 & 55.22 \\
146 & 43.84 & 27.40 & 41.78 & 43.84 & 44.52 & 44.52 \\
64 & 32.81 & 10.94 & 25.00 & 34.38 & 31.25 & 32.81 \\
32 & 34.38 & 28.13 & 34.38 & 37.50 & 34.38 & 34.38 \\
22 & 40.91 & 31.82 & 31.82 & 36.36 & 31.82 & 36.36 \\
42 & 38.10 & 9.52 & 28.57 & 38.10 & 38.10 & 38.10 \\
373 & 42.63 & 23.32 & 38.34 & 42.63 & 41.82 & 42.36 \\
\hline
\end{tabular}
\label{tab:repocoder_testing}
\end{table}

When we compare this to Repo Level Prompting, we can see that the NED relative comparison is comprable with a large 25\% relative improvement to the baseline. You can see this in Table \ref{tab:edit_distance_repo_level_prompting}.

\begin{table}[ht]
\centering
\caption{Normalized Edit Distance (NED) based performance evaluation for repo level prompting.}
\begin{tabular}{lcc}
\hline
\textbf{Method} & \textbf{NED(\%)} & \textbf{Rel. \textuparrow (\%)} \\ \hline
Codex & 30.73 & - \\
RLPG-H & 22.55 & 26.62 \\
RLPG-R & 23.00 & 25.14 \\ \hline
\end{tabular}
\label{tab:edit_distance_repo_level_prompting}
\end{table}

You can also see the impact of each of their different retrieval models. It is worth noting that the file level BM25 is fairly good, but their RLPG-H is the highest performing strategy. Considering how performant the RLPG-BM25 model is, it is possible that taking the prompt generation strategy and doing a BM25 classification may be the optimal option in terms of latency, compute and developer time for many developers who rapidly want something that is highly performant. These results can be seen in \ref{tab:method_comparison_repo_level_prompting}.

\begin{table}[ht]
\centering
\caption{Comparison of methods for repo level prompting.}
\begin{tabular}{lcc}
\hline
\textbf{Method} & \textbf{Success Rate(\%)} & \textbf{Rel. \textuparrow (\%)} \\ \hline
Codex (Chen et al., 2021) & 58.73 & - \\
Oracle & 79.63 & 35.58 \\
Random & 58.13 & -1.02 \\
Random NN & 58.98 & 0.43 \\
File-level BM25 & 63.14 & 7.51 \\
Identifier Usage (Random) & 64.93 & 10.55 \\
Identifier Usage (NN) & 64.91 & 10.52 \\
Fixed Prompt Proposal & 65.78 & 12.00 \\
RLPG-BM25 & 66.41 & 13.07 \\
RLPG-H & 68.51 & 16.65 \\
RLPG-R & 67.80 & 15.44 \\ 
\hline
\end{tabular}
\label{tab:method_comparison_repo_level_prompting}
\end{table}

\begin{figure}[ht]
  \centering
  \includegraphics[width=0.9\linewidth]{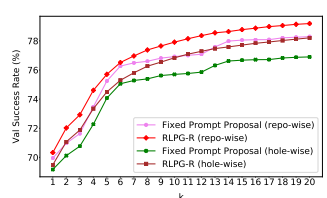}
  \caption{Variation of RLPG and Fixed Prompt Proposal with Num. attempts (k), when averaged over individual repositories (repo-wise) and all holes (hole-wise)}
  \label{fig:perfRepLevel}
\end{figure}

Another experiment the authors of repo level prompting can be seen in Figure \ref{fig:perfRepLevel}. There you can see the impact of choosing the top k prompt proposals the classifiers chose. This shows the steadily increasing performance with further prompts. This is somewhat surprising as you would expect truncation to fit everything into the context window or filling the context window to rapidly be a problem, but performance seems to only rise.

\section{How to build an optimal prompting system}

Considering both of these papers, RepoCoder and Repository Level Prompt Generation can dramatically improve LLM performance when filling gaps in missing chunks of code. On top of that, while one method focuses on iteratively improving generation with retrieval results that are progressively more relevant with an off the shelf retriever, the other focused on building a better retrieval system by using a neural network to figure out exactly what code should be retrieved. It therefore, would make sense to attempt to combine these methods, especially since the sparse retrieval method performed so poorly in Repo Level Prompting. To accomplish this, one would have to train the neural network described in \cite{shrivastava2023repositorylevel} but with the hole filled in and the generated code present. Then to iteratively rerun the generation with the context found. 

\section{Future Work}
On top of naively combining these two methods, another next step would be to integrate them into more complex use cases then doing auto-complete for a single line, function or API request in a repository. One such example of this is described in Software Engineering Bench\cite{jimenez2023swebench}. There they build an extremely robust benchmark where they collect 2,294 github issues that have associated pull requests and test cases. They verify that all of the test cases succeed after the pull request, and if you run the code before the pull request against the test, after the ground truth code is applied to the tests, that some of them fail. This handles several of the edge cases and problems seen in the RepoCoder Test Benchmark. 

In SWE-Bench they provide several additional retrieval systems and tools, but their retrieval system only retrieves entire files which will mostly contain noise and not be useful. Also, many of the issues are very large and must be condensed in some meaningful way. Even their oracle retrieval system pulls in all of the full files that were modified instead of just subsets of them. Because of this the performance of the language models tested is extremely poor, Claude-2 solves 4.8\% and GPT-4 solves 1.7\%. Considering Claude is significantly worse in other benchmarks like HumanEval, 71.2 compared to GPT-4's 86\%\cite{Claude2} \cite{openai2023gpt4}. This is likely because of Claude's long context length and the fact that their retrieval systems only support searching for entire files. Therefore, it likely would be possible to top their online leaderboard at SWE-Bench.com by combining these techniques of iterative generation and custom retrieval proposed in RepoCoder \cite{zhang2023repocoder} and Repo Level Prompting \cite{shrivastava2023repositorylevel} with their benchmark.

The next place to look after all of this may be to agent based LLM usage.\cite{yao2023react}\cite{shinn2023reflexion}\cite{zhou2023language} Building on Reflexion\cite{shinn2023reflexion}, or Language Agent Tree Search \cite{zhou2023language}, the two top performing HumanEval papers which use a mix of reinforcement learning where the LLM is performing the actor and critic work and other methods like monte-carlo tree search to build optimal paths in traversing the problem space. Although these methods are currently computationally expensive and induce high latency, it is likely that as LLM inference time continues to fall that some combination of these methods will create the best performance and will still not force the user to wait for intractably long periods of time.

\section{Conclusion}

This paper has provided a comprehensive review of two innovative approaches for repository-level prompting in Large Language Models: Repository-Level Prompt Generation and RepoCoder. Our analysis revealed that each method has its strengths and trade-offs. RepoCoder's iterative retrieval and generation method excels in refining the prompt process, while Repository-Level Prompt Generation offers a robust framework for evaluating the effectiveness of various prompts against code completion tasks.

We have demonstrated that by leveraging these methods, LLMs can be prompted to produce contextually relevant and syntactically coherent code completions that align with the complex requirements of real-world software repositories. The iterative nature of RepoCoder aligns well with the dynamic development environments, offering incremental improvements in code generation. Conversely, the classification-based approach of Repository-Level Prompt Generation provides a more direct path to identifying effective prompts, although it may require more extensive training data.

Our comparative analysis underscores the potential for integrating these strategies to harness their combined strengths. By doing so, we can address the current limitations of sparse retrieval methods and pave the way for more sophisticated code generation models. The promising results from applying these methods suggest that we are on the cusp of a new era in automated coding, where LLMs can not only assist in code generation but also evolve with the changing landscapes of software development.

As we look to the future, the integration of these methods into more complex systems, such as those described in Software Engineering Bench, represents the next frontier for LLMs. The potential for topping performance leaderboards by combining iterative generation with custom retrieval systems is substantial and warrants further exploration.

Ultimately, the journey towards fully autonomous and intelligent coding assistants is just beginning. We anticipate that as LLM inference times decrease, the amalgamation of agent-based LLM usage, reinforcement learning, and other advanced techniques will set new benchmarks in the field. The research community is encouraged to build upon the insights presented in this study to create LLMs that not only augment the coding process but also contribute to the art of programming itself.

\pagebreak
\bibliographystyle{aaai} 
\bibliography{bib}

\begin{thebibliography}{}

\bibitem[\protect\citeauthoryear{Amati}{2009}]{Amati2009}
Amati, G.
\newblock 2009.
\newblock {\em BM25}.
\newblock Boston, MA: Springer US.
\newblock  257--260.

\bibitem[\protect\citeauthoryear{Anthropic}{2023}]{Claude2}
Anthropic.
\newblock 2023.
\newblock Claude-2 technical report.

\bibitem[\protect\citeauthoryear{Chen \bgroup et al\mbox.\egroup }{2021}]{chen2021codex}
Chen, M.; Tworek, J.; Jun, H.; Yuan, Q.; de~Oliveira~Pinto, H.~P.; Kaplan, J.; Edwards, H.; Burda, Y.; Joseph, N.; Brockman, G.; Ray, A.; Puri, R.; Krueger, G.; Petrov, M.; Khlaaf, H.; Sastry, G.; Mishkin, P.; Chan, B.; Gray, S.; Ryder, N.; Pavlov, M.; Power, A.; Kaiser, L.; Bavarian, M.; Winter, C.; Tillet, P.; Such, F.~P.; Cummings, D.; Plappert, M.; Chantzis, F.; Barnes, E.; Herbert-Voss, A.; Guss, W.~H.; Nichol, A.; Paino, A.; Tezak, N.; Tang, J.; Babuschkin, I.; Balaji, S.; Jain, S.; Saunders, W.; Hesse, C.; Carr, A.~N.; Leike, J.; Achiam, J.; Misra, V.; Morikawa, E.; Radford, A.; Knight, M.; Brundage, M.; Murati, M.; Mayer, K.; Welinder, P.; McGrew, B.; Amodei, D.; McCandlish, S.; Sutskever, I.; and Zaremba, W.
\newblock 2021.
\newblock Evaluating large language models trained on code.

\bibitem[\protect\citeauthoryear{Jimenez \bgroup et al\mbox.\egroup }{2023}]{jimenez2023swebench}
Jimenez, C.~E.; Yang, J.; Wettig, A.; Yao, S.; Pei, K.; Press, O.; and Narasimhan, K.
\newblock 2023.
\newblock Swe-bench: Can language models resolve real-world github issues?

\bibitem[\protect\citeauthoryear{Levenshtein}{1965}]{Levenshtein1965BinaryCC}
Levenshtein, V.~I.
\newblock 1965.
\newblock Binary codes capable of correcting deletions, insertions, and reversals.
\newblock {\em Soviet physics. Doklady} 10:707--710.

\bibitem[\protect\citeauthoryear{OpenAI}{2023}]{openai2023gpt4}
OpenAI.
\newblock 2023.
\newblock Gpt-4 technical report.

\bibitem[\protect\citeauthoryear{Shinn \bgroup et al\mbox.\egroup }{2023}]{shinn2023reflexion}
Shinn, N.; Cassano, F.; Berman, E.; Gopinath, A.; Narasimhan, K.; and Yao, S.
\newblock 2023.
\newblock Reflexion: Language agents with verbal reinforcement learning.

\bibitem[\protect\citeauthoryear{Shrivastava, Larochelle, and Tarlow}{2023}]{shrivastava2023repositorylevel}
Shrivastava, D.; Larochelle, H.; and Tarlow, D.
\newblock 2023.
\newblock Repository-level prompt generation for large language models of code.

\bibitem[\protect\citeauthoryear{Yao \bgroup et al\mbox.\egroup }{2023}]{yao2023react}
Yao, S.; Zhao, J.; Yu, D.; Du, N.; Shafran, I.; Narasimhan, K.; and Cao, Y.
\newblock 2023.
\newblock React: Synergizing reasoning and acting in language models.

\bibitem[\protect\citeauthoryear{Zhang \bgroup et al\mbox.\egroup }{2023}]{zhang2023repocoder}
Zhang, F.; Chen, B.; Zhang, Y.; Keung, J.; Liu, J.; Zan, D.; Mao, Y.; Lou, J.-G.; and Chen, W.
\newblock 2023.
\newblock Repocoder: Repository-level code completion through iterative retrieval and generation.

\bibitem[\protect\citeauthoryear{Zhou \bgroup et al\mbox.\egroup }{2023}]{zhou2023language}
Zhou, A.; Yan, K.; Shlapentokh-Rothman, M.; Wang, H.; and Wang, Y.-X.
\newblock 2023.
\newblock Language agent tree search unifies reasoning acting and planning in language models.

\end{thebibliography}

\end{document}